\documentclass[%
 aip,
cp,  
 amsmath,amssymb,
 reprint,%
]{revtex4-2}

\usepackage{graphicx}
\usepackage{dcolumn}
\usepackage{bm}
\usepackage[mathlines]{lineno}

\usepackage[utf8]{inputenc}
\usepackage[T1]{fontenc}
\usepackage{xcolor}
\usepackage{hyperref}  
\usepackage{cleveref} 

\usepackage{mathptmx}

\newcommand{\uB}{{MicroBooNE~}}
\newcommand{\agd}{\mbox{ALPHAGAZ\TM-2}}
\newcommand{\agu}{\mbox{ALPHAGAZ\TM-1}}
\newcommand{\TM}{\texttrademark}

\begin{document}

\title{Measurement of the Argon Purity by ICP-MS and Results of the Analysis of the Gas Used for the MicroBooNE Experiment}

\author{R.~Santorelli} 
 \email[Corresponding author: ]{roberto.santorelli@ciemat.es}
\author{A.I.~Barrado~Olmedo}%
\author{E.~Conde~Vilda}%
\author{M.~Fernandez~Diaz}%
\author{V.~Pesudo}%
\author{L.~Romero}%
 \affiliation{
 Centro de Investigaciones Energ\'{e}ticas, Medioambientales y Tecnol\'{o}gicas (CIEMAT), Av.~Complutense~40, 28040 Madrid, Spain
}

\date{\today} 

\begin{abstract}
Measuring the argon purity is critical for all Ar-based rare event research experiments. Mass spectrometry is typically used to assess the uranium and thorium contamination in samples of the materials used to build low-background detectors; however, this technique has the potential to provide other valuable information that is currently not exploited. We have shown that, by \mbox{ICP-MS}, it is possible to identify and quantify common chemical contaminants in argon. Preliminary tests were done with the gas extracted from the experiments MicroBooNE at FNAL and ArDM at LSC. In the former case, we evidenced relevant nitrogen contamination well above the one measured in the commercial argon gas. In ArDM, we identified and quantified the presence of mercury in the gas used for its science run. 
In both cases,  the presence of krypton ($\sim$ppb) and xenon ($\sim$10s ppb) in argon gas has been established. 

\end{abstract}

\maketitle

\section{\label{sec:Intro}\protect Introduction}

The efficient free electron collection and photon detection are the two fundamental features that make liquid argon (LAr) viable as an active medium in rare event search experiments. 
Small concentrations of impurities (typically O$_2$, N$_2$, H$_2$O, and CO+CO$_2$) at the level $\lesssim$ 1 ppm are usually present in commercially available LAr (best grade) due to the industrial air separation process. These impurities can affect the detector performance as they significantly reduce the charge and light signals by ionization electron attachment and scintillation light quenching \cite{WArP:2008rgv,WArP:2008dyo}. 
Purification systems (oxygen reagents, getters, and molecular sieves) can be effective in reducing the O$_2$, H$_2$O, and CO+CO$_{2}$ contamination \cite{Antonello:2014eha}, however, the next generation of argon experiments requires extremely high LAr purity ($\ll$ 1 ppb)   with many meters of electron drift \cite{DUNE:2017pqt}. At the same time, methods to remove the residual N$_2$ content in argon are, to a large extent, more difficult than the ones required for the other contaminants. 
The evidence of charge or light signal quenching in the detector is the sign of a possible air leak, insufficient cleaning of the detector, or contamination of the argon before filling. Only using the light and charge detector's data, it is tough to determine whether the reduced signal is due to one or more chemical impurities and identify the specific contaminants.

On the other hand, inductively coupled plasma mass spectroscopy (ICP-MS) is
a robust and widespread technique used to determine the concentrations in the trace and ultra-trace range by counting the number of atoms of the elements which are detected \cite{Laubenstein:2020rbe,Dobson:2017esw}. The nebulized solution or the ablated material is transported by Ar as carrier gas into the inductively coupled plasma ion source of the sensitive double-focusing sector. The plasma source ionizes the sample, and then the ion beam is transferred to a quadrupole m/z filter. With this technique, concentrations of the order of ppq (pg/L) can be obtained, limited mainly by the procedure required for the sample preparation before the injection into the plasma. Both elemental and isotopic information can be obtained in a relatively short measuring time ($\approx$~min). 

The stringent requirements of the new generation of rare-event research experiments in terms of background minimization have made the ICP-MS technique very popular for measuring the contamination of the upper decay chains of U and Th in the detector materials. 
Particularly, mass spectroscopy makes it possible to determine the U and Th concentration in the sample with better sensitivity than the one achieved with gamma ray spectroscopy using high-purity germanium crystals.

An important limit in the trace, ultra-trace, and isotope ratio measurements by ICP-MS is the occurrence of a multitude of different interferences in mass spectra given by isobaric singly charged atomic ions (e.g., $^{238}$Pu$^{+}$,  $^{238}$U$^{+}$), doubly charged atomic ions at the same nominal mass (e.g., $^{90}$Sr$^{+}$, $^{180}$Hf$^{++}$), or by the formation of polyatomic ions (e.g., $^{238}$UH$^{+}$, $^{40}$Ar$^{14}$N$^{+}$) \cite{BECKER2012833}. 
The variety of polyatomic ions is larger at lower masses, thus, it is somewhat tricky to measure the contamination of elements like O$_{2}$ and N$_{2}$  with this technique. 
A typical way to remove the spurious signal given by polyatomic ions is by using non-reactive gases (e.g., He) in a collision cell. A process called kinetic energy discrimination (KED) can attenuate all polyatomic interferences based on their size. KED exploits the fact that all polyatomic ions are larger than analyte ions of the same mass, so they collide with the cell gas more often as they pass through the cell, emerging with lower residual energy. These low-energy ions are excluded from the ion beam by a bias voltage at the cell exit.

A relevant aspect of this technique is that the ICP mass spectrometers use argon  to generate the plasma and as a gaseous carrier due to its relatively low cost and the high first ionization potential, which ensures that the sample to be measured remains ionized.  Thus, it is possible to inject Ar gas into the mass analyzer without adding any sample, directly analyzing the gas and providing the m/z spectrum of the contaminants in the gas. A few grams of Ar could be sufficient for this measurement.  
This procedure doesn't require the chemical sample preparation necessary for the digestion of solid materials. It is relatively simple and has the potential to provide essential data on the chemical contaminants in the argon,  something that the rare event search community does not generally exploit within the \mbox{ICP-MS} technique. Eventually, it is possible to identify the chemical elements responsible for the charge or light signals quenching in an argon-based experiment. 

The CIEMAT-DM group performed preliminary tests with the gas extracted from the ArDM experiment at the Canfranc Underground Laboratory  (LSC) in Spain, demonstrating the feasibility of this technique \cite{ArDM:2016jbw}. The ArDM gas, nominally of the \agd~type (N60, 99.9999\% purity), was screened and also compared to \agu~ (N50, 99.999\% purity) supplied by Air Liquide Spain to CIEMAT-Madrid \cite{airl}.
Data with different gas flows were taken by changing the input pressure of the gas. The results evidenced traces of Xe and Kr in the argon with an estimated concentration of less than 1\,ppb for the ArDM gas and at least one order of magnitude more for \agu. The only other trace evidence, which is by far the most dominant \cite{ArDM:2016jbw}, has been identified as Hg at the level of 10\,$\pm$\,5\,ppb. The Hg identification was surprising but is clearly established by the relative natural abundance of the different Hg isotopes. The Hg signal has been calibrated with standard samples of known concentrations (Standard TraceCERT\textsuperscript{\textregistered} from Sigma-Aldrich, 1000\,mg/L Hg with 12\% HNO$_3$), and the error on the measured contamination is mainly given by the uncertainty determining the argon flux through the instrument.  
It has been found that the Hg concentration is three times larger in the ArDM gas compared to \agu. A possible explanation was given by the admixture of H$_{2}$ gas in the argon production process in order to reduce the O$_{2}$ content of natural argon. The water produced in this process is later removed by a drier system. The manufacturer confirmed that the H$_{2}$ used during the process is obtained by electrolysis with Hg electrodes, so an unknown level of contamination of Hg vapor can be assumed for the H$_{2}$ gas. The higher Hg contamination found for the ArDM gas (\agd) in respect to the \agu, could originate in the heavier treatment of the argon with H$_{2}$ gas necessary to reduce the O$_{2}$ contamination further. 
This unexpected contamination had to be accounted for in the light propagation model of the ArDM experiment.

\begin{figure} [t!]
\includegraphics[width=0.7\textwidth]{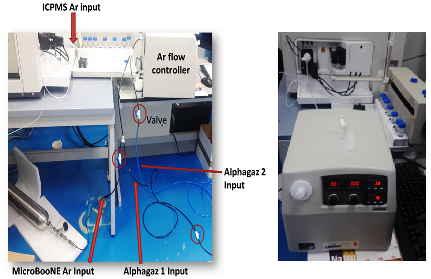}
\caption{\label{fig:Setup} Experimental setup with the three argon lines, valves and the mass spectrometer (Left). Ar flow controller and ICP-MS gas inlet to the injectors to the plasma region (Right).}
\end{figure}

A more recent analysis was performed with the gas extracted from the MicroBooNE detector. Here we identified some typical argon contaminants and compared the ICP-MS results with commercially available argon gas. This study opens new perspectives for the testing and monitoring the purity of argon gas used for future massive Ar-based experiments.

\section{\label{sec:Setup}\protect Experimental setup}

A sample of the argon gas was extracted from the \uB detector at FNAL and shipped to CIEMAT in a bottle at 2.5 bar pressure (12 standard liters). The mass spectrum of the \uB gas has been compared with the spectra obtained analyzing commercial gas \agu~ and \agd~ from Air Liquide Spain.
In order to introduce the gas from the different Ar bottles, a gas line splitter with three valves is used. The gas from one bottle is injected into the plasma with the appropriate opening and closing of the valves without disconnecting the input of the line. In this way, it is possible to ensure the same analysis conditions, avoiding any manipulations that could affect the plasma when changing the gas sample. The Ar inlet flow rate is controlled with a mass flow meter installed within a desolvating nebulizer system (CETAC Aridus I, Fig. \ref{fig:Setup}-right). 
A quadrupole-based ICP-MS iCAPQ\TM\ (Thermo Scientific) with a discrete dynode SEM detector operated in both digital and analog modes is used in our experiment. To introduce the Ar gas to be analyzed into the mass spectrometer, the ICP-MS  inlet system was slightly modified, making the Ar gas go directly to the center of the plasma through the injector. 
Data with the three argon samples are taken in the same pressure and temperature conditions, keeping a constant gas flow of 3.5 SLPM. The  software Q-tegra\TM\ Intelligent Scientific  Data Solution\TM\ (ISDS) from Thermo Scientific converts the  electric signal from the ICP-MS into intensity signal, expressed in counts per second (cps) for each mass unit value, presenting the results in a graphical way.

To ensure that the same instrumental background is maintained during the measurement of the different samples, the ICP-MS input line is flushed for 30 s before data are recorded. 
Multiple sets of data were acquired by switching from one bottle to another. Different runs taken for the same sample are statistically consistent, thus demonstrating the correct functioning of the valve system and the stability of the measurement conditions during the experiment.

\begin{figure} [t!]
\includegraphics[width=0.65\textwidth]{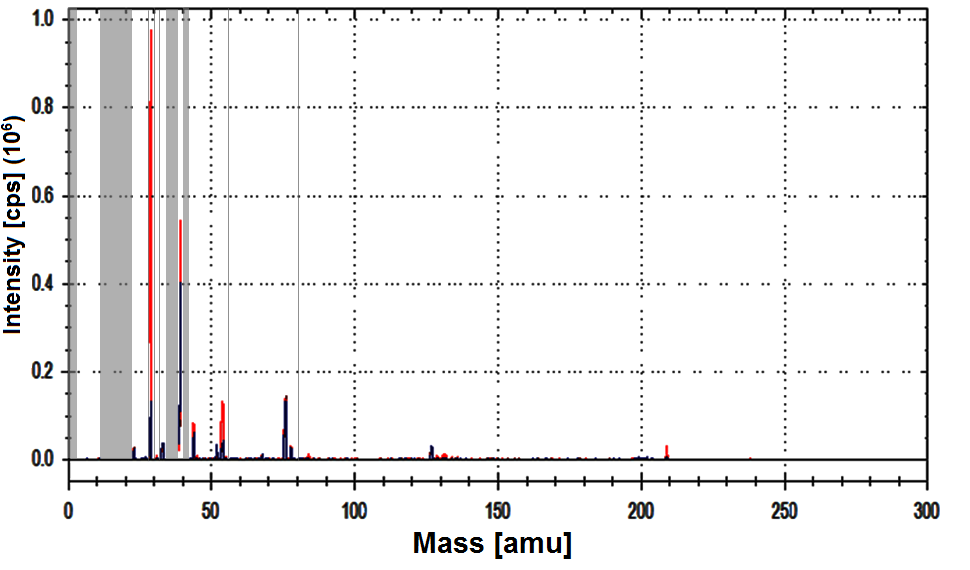}
\caption{\label{fig:Full} Superposition of the full m/z spectra taken with the \uB gas (red) and commercial gas from Air Liquide (\agd, blue). The gray lines correspond to the values of the mass not measured by the instrument.}
\end{figure}

\section{\label{sec:MB}\protect Analysis of the argon used in the MicroBooNE experiment}

The m/z spectra recorded with the ICP-MS for the three samples are shown in Fig. \ref{fig:Full} in the full atomic mass units (amu) range. The peaks with amu < 60 are the most prominent in terms of the total number of counts per second; however, it is worth stressing that the efficiency of the instrument is not linear and depends on the particular isotope. Calibrated samples with known concentrations of a specific element would be needed in order to convert precisely the absolute cps to concentration in argon.
We can divide the spectra into three sub-regions (amu < 60,  60 < amu < 150, and amu > 150), discussing the differences in the contamination of the three samples mainly through a comparative analysis.

\subsection{\label{sec:ML}\protect m/z < 60 amu}

\begin{figure} [t!]
\includegraphics[width=0.45\textwidth]{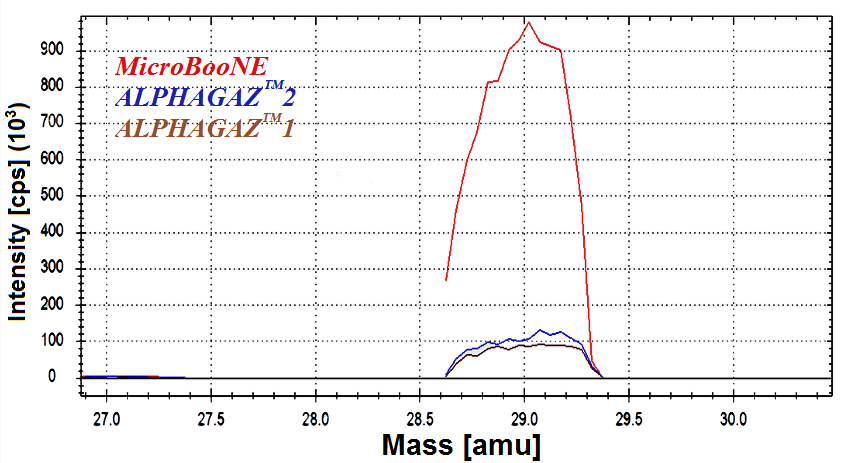}
\includegraphics[width=0.45\textwidth]{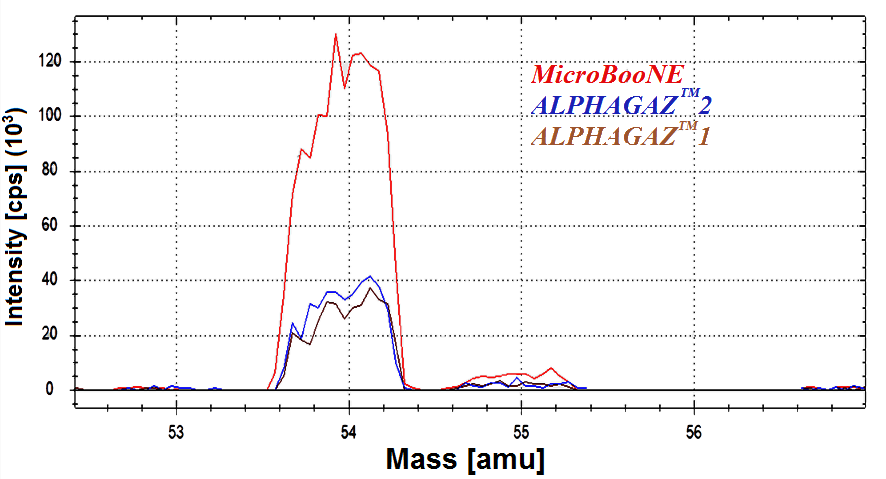}
\caption{\label{fig:ML} Mass spectrometry analysis of the MicroBooNE gas (red) compared with the results of the commercial Air Liquide gas (blue and brown). The peaks in the region m/z < 60 are explained as N$_{2}$ contamination in Ar. }
\end{figure}

In the low-mass region, peaks above the background, defined as the counts per second measured with no argon flow, are evident for mass units 29, 54, and 55 (Fig. \ref{fig:ML}). In the first case, a possible explanation for this peak can be given by the presence of  $^{29}$Si isotope (4.7\% of natural silicon) in the argon sample. However, this hypothesis is disfavoured by the absence of the peaks at 28 and 30, which should be detected in case of silicon contamination  ($^{28}$Si 92.2\% natural abundance (NA) and  $^{30}$Si 3.1\% NA). The hypothesis of contamination of just one silicon isotope is unlikely and can be discarded. 
On the contrary, we explain the peak at 29 amu as the formation of polyatomic ions $^{14}$N$^{15}$N given by the two naturally-occurring isotopes of nitrogen. A simple confirmation of this nitrogen contamination in the \uB gas could be provided by the contemporary detection of peaks at 14,15 or 28 amu, corresponding to the species $^{14}$N, $^{15}$N, and $^{14}$N$^{14}$N, however, these values are not measured during the scanning performed by the iCAPQ\TM\ instrument to avoid the possible saturation of the detector.

At the same time,  the presence of nitrogen in the gas can be confirmed by the detection of the other two peaks at 54 and 55 mass units. Rather than interpreting these peaks as the unlikely contamination of specific isotopes of some transition metals (Cr, Mn, Fe, etc.), a more natural explanation can be provided by the formation of polyatomic ions given by the combination of the two isotopes of nitrogen with $^{40}$Ar. The solid confirmation of this hypothesis is given by the comparison of the relative intensity of the peaks at 54 and 55 mass units, which is in the order of the one expected from the natural abundance of the two isotopes in natural nitrogen (99.4\% and 0.4\% respectively for $^{14}$N and $^{15}$N).
It is important to evidence that nitrogen is detected in the three samples of gas, with significantly larger contamination in the \uB sample. The direct quantification of this contamination in argon gas would have required specific ICP-MS characterization with calibrated samples and was not done. However, nitrogen in the \uB gas of the order of $\sim$ppm can be estimated based on the purity specifications of the \agd~ from Air Liquide ($\lesssim$ ppm for N$_{2}$).

\subsection{\label{sec:MI}\protect m/z > 60 amu and m/z < 150 amu}

\begin{figure} [t!]
\includegraphics[width=0.45\textwidth]{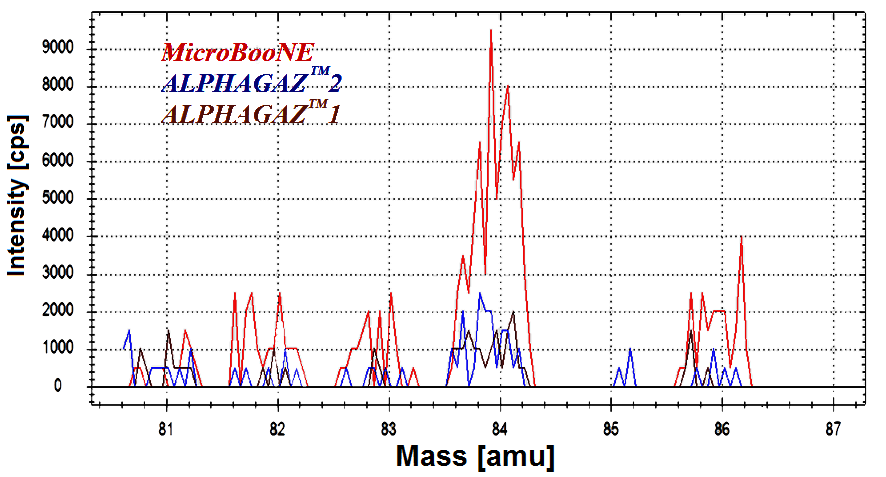}
\includegraphics[width=0.45\textwidth]{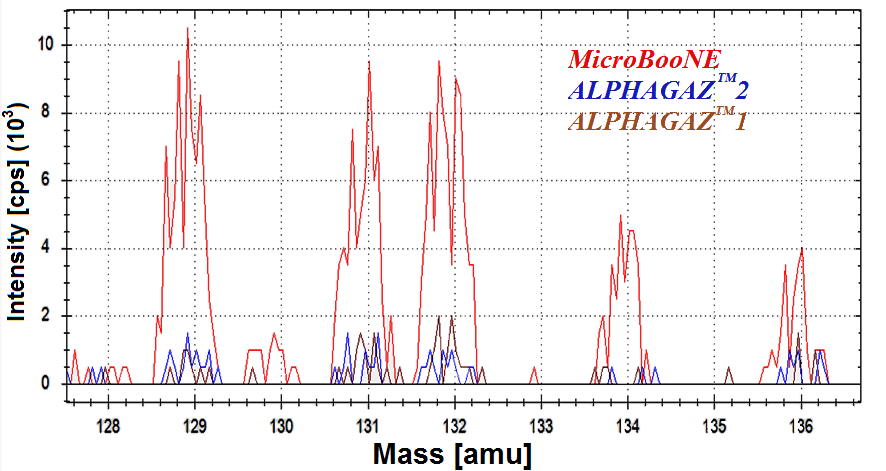}
\caption{\label{fig:MM} m/z region [60,150] amu: the peaks in this region can be explained as Kr (left) and Xe (right) contamination.}
\end{figure}

The intensity of the peaks detected in the range [82,86] amu is shown in Fig. \ref{fig:MM}-left, proving a larger contamination in the \uB sample. The interpretation of these peaks in terms of krypton contamination is supported by the contemporary detection of all the expected isotopes, with a relative intensity in agreement with the natural abundance of the Kr. 
A similar conclusion is valid for the peaks in the [128,135] mass units region (Fig. \ref{fig:MM}-right), which is consistent with the detection of xenon. The presence of both Kr and Xe in our samples is possible since argon is distilled from the atmosphere and some residual contamination by other noble gases is plausible. Also in this case, the contamination of the \uB gas is significantly larger than that found in the other two commercial samples. 

An estimate of the Kr and Xe contamination based on a generic calibration curve of the ICP-MS in this region puts the value in the 1-10 ppb range.

\subsection{\label{sec:MH}\protect m/z > 150 amu}

\begin{figure} [t!]
\includegraphics[width=0.45\textwidth]{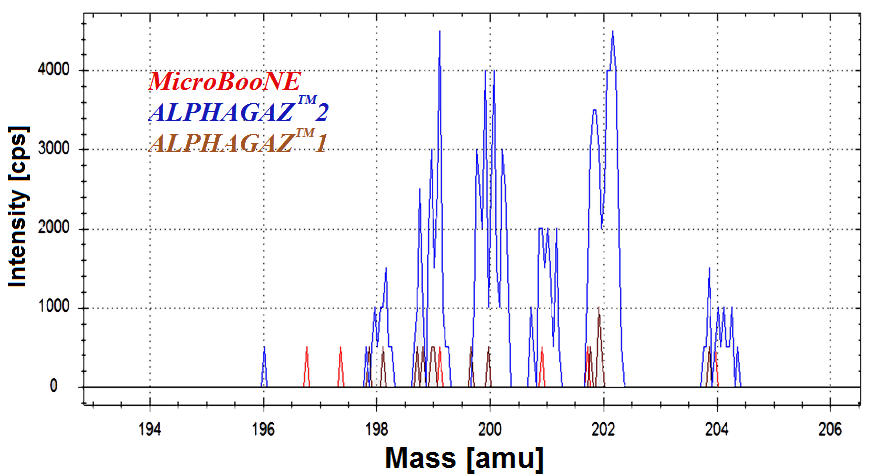}
\caption{\label{fig:MH} Mass spectrometry analysis of the MicroBooNE gas in the Hg region: no evident Hg contamination is found in the \uB gas. }
\end{figure}

In the spectrum region with masses above xenon, peaks are found only for mass unit values consistent with Hg contamination (Fig. \ref{fig:MH}). As in the previous cases, the unambiguous identification of the mercury contamination can be provided by the simultaneous detection of the Hg isotopes with the relative intensity of the m/z peaks that agrees with their natural abundance. An evident Hg contamination is found only in the \agd~ sample, while the intensity of the peaks in the \uB sample is consistent with the background. This result confirms that the Hg contamination is related to the specific production process of the Ar in the Spanish plants \cite{ArDM:2016jbw}. 

\section{\label{sec:ked}\protect Investigation of the N$_{2}$ contamination with the collision cell}

\begin{figure} [t!]
\includegraphics[width=0.55\textwidth]{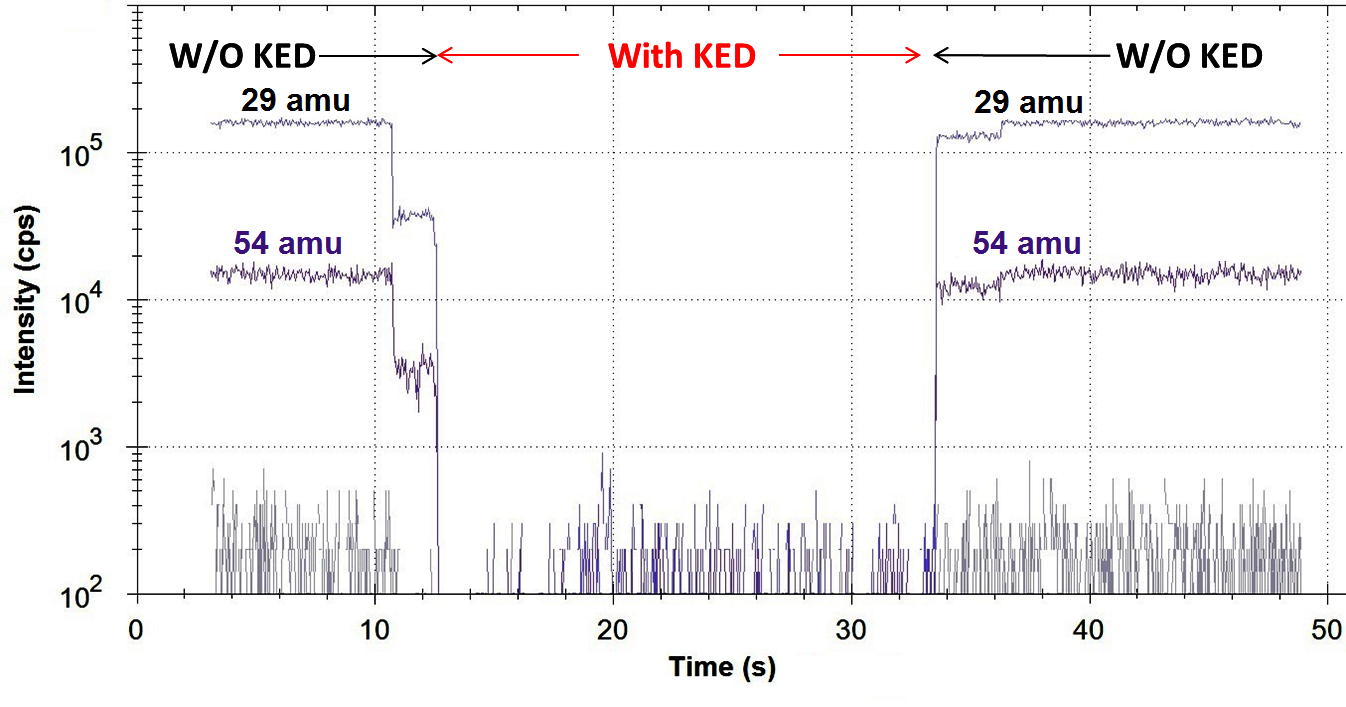}
\caption{\label{fig:ked} Variation of the counts per second possibly produced by the N$_{2}$ contamination in the \uB argon with and without the collision cell. }
\end{figure}

In order to confirm the interpretation of the peaks at m/z = 29, 54, and 55 amu in terms of N$_{2}$ contamination, a preliminary test has been carried out using a kinetic collision cell mode (CCT-KED) operated with He gas. Polyatomic ions have larger collision cross-sections, undergo more collisions, and lose more kinetic energy than the atomic ions with the same m/z value. While the atomic ions surmount the energy barrier placed downstream of the cell, the polyatomic ions do not; thus, a reduction of the intensity of the peaks they produce is expected. 
In Fig. \ref{fig:ked}, the variation of the counts per second of the peaks at m/z= 29 and 54 for the \uB argon is shown with and without the collision cell. The N$_{2}$ produced peaks disappear with the CCT-KED, and the counts per second are compatible with the instrument's background in these conditions. For comparison, the intensity of the peaks produced by the Kr and Xe contaminations are just reduced by a factor 2-3 with the collision cell in operation, thus proving the effective rejection of the polyatomic ions.

\section{Conclusion}
We propose a new method to investigate argon purity based on the ICP-MS technique. This procedure doesn't require the chemical sample preparation necessary for the digestion of solid materials; it is relatively simple and only requires a few grams of Ar. The m/z spectrum of the argon obtained with the ICP-MS analysis has the potential to identify the chemical contaminants which can be responsible for the charge or light signals quenching in an argon-based experiment. Results can be obtained in a time scale of $\sim$ min, making it possible to detect any tiny leak in the system promptly. 
The viability of this technique has been recently proved by the analysis of a sample of argon extracted from the \uB detector at FNAL. We compared the m/z spectrum of this gas with the ones from commercial argon with different purity grades, finding relevant nitrogen contamination at the ppm level in the \uB gas. In all the samples, Xe and Kr contamination at the level of ppb has been undoubtedly identified by identifying all the natural isotopes of those elements. 
We plan to study the possible application of this technique for real-time monitoring of gas purity for the future generation of argon detectors. 

\begin{acknowledgments}
This research is funded by the Spanish Ministry of Science (MICINN) through the grant PID2019-109374GB-I00 and supported by the “Física de partículas” Unit of CIEMAT through the grant MDM-2015-0509. V. Pesudo is funded by the "Atraccion de Talento" grant 2018-T2/TIC-10494. We would like to thank the MicroBooNE collaboration, Fermilab operations staff, and Fermilab cryogenic engineering staff for providing the argon sample and collaborating on the interpretation of this data.
We acknowledge the techincal support of Javier Lacal from thermo-Fisher España.

\end{acknowledgments}

\nocite{*}
\bibliography{aipsamp}

\end{document}